\documentstyle[equations,epsf,12pt]{article}
\begin{document}

\begin{flushright}
IPNO/TH 97-06
\end{flushright}
\vspace{1 cm}
\begin{center}
{\large \textbf{The infinite mass limit of the two-particle Green's 
function \protect \\ in QED}}
\vspace{1.5 cm}

H. Jallouli\footnote{E-mail: jallouli@ipno.in2p3.fr .}
and H. Sazdjian\footnote{E-mail: sazdjian@ipno.in2p3.fr .}\\
\renewcommand{\thefootnote}{\fnsymbol{footnote}}
\textit{Division de Physique Th\'eorique\footnote{Unit\'e de Recherche des
Universit\'es Paris 11 et Paris 6 associ\'ee au CNRS.},
Institut de Physique Nucl\'eaire,\\
Universit\'e Paris XI,\\
F-91406 Orsay Cedex, France}
\end{center}
\vspace{1.5 cm}

\begin{center}
{\large Abstract}
\end{center}

The behavior of the two-particle Green's function in QED is analyzed in 
the limit when one of the particles becomes infinitely massive. 
It is found that the dependences of the Green's function on the relative 
times of the ingoing and outgoing particles factorize and that the 
bound state spectrum is the same as that of the Dirac equation with the 
static potential created by the heavy particle. The Bethe--Salpeter wave 
function is also determined in terms of the Dirac wave function. The 
present result excludes the existence, in the above limit, of abnormal 
solutions due to relative time excitations as predicted by the 
Bethe--Salpeter equation in the ladder approximation.

\newpage

\section{Introduction}

It is generally admitted that the infinite mass limit of the two-particle
Green's function in QED (or in similar abelian theories) yields for the
lighter particle the Dirac equation in the presence of the static 
potential created by the heavy one. The corresponding derivations are
essentially  based on mass-shell \cite{b,d,ng} or quasi-mass-shell
\cite{g,fh,ms} conditions. On the other hand, it is known that the 
Bethe--Salpeter equation in the ladder approximation leads to the appearance
of ``abnormal'' solutions in the spectrum, due to the relative time
excitations \cite{w,ck,n1,n2}. Taking the infinite mass limit (for one of 
the particles) does not remove from the spectrum the above solutions
\cite{ck} and therefore the corresponding limiting equation is not
equivalent to the Dirac equation with a static potential. It is 
noteworthy to mention that in 
the static limit, when both particle masses go to infinity, the abnormal
solutions still persist in the ladder approximation \cite{otu}.
\par
In a recent work, we showed that in the static limit
the two-particle Green's function can exactly be calculated and the 
corresponding bound state spectrum does not display any abnormal solution
\cite{js}. This result strongly suggests that the appearance of the
abnormal solutions is rather a consequence of the ladder approximation
(with covariant propagators) of the Bethe--Salpeter equation. It is then 
natural to expect that also the same conclusion might hold in the 
one-particle limit.
\par
The purpose of the present article is to study the (one-particle)
infinite mass limit of the two-particle Green's function without making
any mass-shell or quasi-mass-shell assumptions on the heavy particle
and therefore not constraining the relative time evolution of the 
system. We shall show that the dependences of the Green's function
on the relative times of the ingoing and outgoing particles factorize
and that its exact spectrum in the above limit (in the absence of 
radiative corrections) is given by that
of the Dirac equation of the lighter particle in the presence of the 
static potential. The relative time variables, without being absent 
from the expressions of the Green's function and of the Bethe--Salpeter
wave function do not induce energy excitations, but rather ensure by
their presence the correct spectral properties as dictated by quantum
field theory \cite{w,n1}. This result also also allows the determination of 
the Bethe--Salpeter wave function in terms of the Dirac wave function.
\par
While in the above calculations radiative corrections of the lighter
particle are neglected (or partly taken into account in covariant gauges), 
this approximation does not seem crucial for the
derivation of the factorization property of the Green's function in the
relative time variables, since this feature is rather related to the heavy
particle properties. By appropriately generalizing the static potential so as 
to include the full radiative corrections of the lighter particle,
one should still be able to prove the main qualitative results as described
above.
\par
The QED bound state spectrum being gauge invariant \cite{by,ffh} it is 
actually sufficient to prove the above properties in one particular gauge. 
We shall choose the Coulomb gauge for a detailed derivation, but shall 
also briefly sketch the line of calculation with linear covariant gauges.
\par

\section{The two-particle Green's function with an infinite mass}
\setcounter{equation}{0}

Our starting point is the QED lagrangian density for two fermions with 
masses $m_1$ and $m_2$ and charges $e_1=e_2=-e$:
\begin{eqnarray} \label{e1}
\mathcal{L}&=&\overline\psi _1  (i\gamma . \partial  - m_1 + e \gamma . A )
\psi _1 +
\overline\psi _2  (i\gamma . \partial  - m_2 + e \gamma . A ) \psi _2
\nonumber \\
& &\ \ \ \ \ \ \ \ \ \ \ -
\frac{1}{4} F_{\mu\nu}F^{\mu\nu} + \mathcal{L}_{g.f.} ,
\end{eqnarray}
where $\mathcal{L}_{g.f.}$ is the gauge fixing part of the lagrangian
density. The two-particle Green's function is defined by the functional
integral:
\begin{eqnarray}  \label{e2}
G(x_1,x_2,x_1',x_2')&=& \mathcal{N}\int D\psi D\overline\psi DA
\psi_1(x_1) \psi_2(x_2')
\overline \psi_2(x_2) \overline\psi_1(x_1')
e^{iS(\psi,\overline\psi,A)} \nonumber  \\
&\equiv&<\psi_1(x_1) \psi_2(x_2')
\overline\psi_2(x_2) \overline\psi_1(x_1')
e^{iS(\psi,\overline\psi,A)} >  ,
\end{eqnarray}
where $S=\int d^4x \mathcal{L}(x)$ is the action. We shall be interested 
by the bound state spectrum of the system fermion 1-antifermion 2.
We consider particle 2 as the heavy one and take the limit of large 
$m_2$. This limit can be studied by treating perturbatively the (covariant) 
kinetic energy term of particle 2. Such a situation has often been met in 
heavy quark effective field theories \cite{hq}. A way of implementing 
this limit is to make successive transformations on the fermion 2 field, 
the first of which being \cite{kt}:
\begin{equation} \label{e3}
\psi_2\rightarrow \exp\left((i\gamma_i\partial^i+e\gamma_iA^i)/(2m_2)
\right) \psi_2, \ \ \ \ \ \overline \psi_2 \rightarrow \overline 
\psi_2 \exp\left((-i\gamma_i\stackrel{\leftarrow}{\partial^i}+e\gamma_i
A^i)/(2m_2)\right).
\end{equation}
This transformation has been shown to have unit jacobian \cite{kt}. 
It eliminates at order $m_2^0$ the (covariant) kinetic energy part of 
the fermion 2 lagrangian density and the corresponding free propagator
becomes the static one:
\begin{eqnarray} \label{e4}
G_{20}(x_2'-x_2) &=& \frac{1}{2}(1+\gamma_0) e^{-im_2(x_2^{\prime 0}
-x_2^0)}\theta (x_2^{\prime 0}-x_2^0)\delta^3(\mathbf{x}_2'-\mathbf{x}_2)
\nonumber \\
& & +\frac{1}{2}(1-\gamma_0) e^{-im_2(x_2^0-x_2^{\prime 0})}\theta (x_2^0
-x_2^{\prime 0})\delta^3(\mathbf{x}_2-\mathbf{x}_2').
\end{eqnarray}
The antifermion contribution is selected when $x_2^0>x_2^{\prime 0}$
and we shall henceforth consider only this case ($\gamma_0$ has then the
eigenvalue $-1$).
\par
Integration of the functional integral  (\ref{e2}) with respect to the 
photon field amounts to replacing the action $S$ by an effective one 
$\overline S = \int d^4 x d^4 y\overline {\cal L}(x,y)$, with
\begin{eqnarray} \label{e5}
\overline {\cal L}(x,y) &=& \left(\overline \psi_1(i\gamma .
\partial -m_1)\psi_1 + \psi_2(i\gamma_0\partial^0-m_2)\psi_2 
\right)_x\delta^4(x-y)\nonumber \\
& & +\frac{ie^2}{2}\Big(j_{1\mu}(x)+j_{2\mu}(x)\Big) D^{\mu\nu}(x-y)
\Big(j_{1\nu}(y)+j_{2\nu}(y)\Big),
\end{eqnarray}
where $j_{a\mu}\equiv\overline \psi_a\gamma_{\mu}\psi_a$, $a=1,2$, is the 
current of particle $a$, with $j_{2\mu}=g_{\mu}^{\ 0} j_{20}$, and 
$D_{\mu\nu}$
is the photon propagator. We shall neglect in the following tha radiative 
corrections and therefore shall maintain in the interaction part of 
$\overline {\cal L}$ the crossed terms $j_1Dj_2 +j_2Dj_1$ only.
(Actually, the radiative corrections relative to the heavy particle 2 can 
explicitly be calculated \cite{js} and the latter approximation concerns 
rather the lighter particle 1.) 
\par
Use of Wick's theorem with respect to particle 2 together with formula
(\ref{e4}) and condition $x_2^0>x_2^{\prime 0}$ yields for the
Green's function (\ref{e2}) the expression:
\begin{eqnarray}
\label{e6}
G(x_1,x_2,x_1',x_2') &=& G_{20}(x_2'-x_2) <\psi_1(x_1)\overline
\psi_1(x_1') \nonumber \\
&\times& \exp\bigg\{i\int d^4y_1 \overline\psi_1(y_1)\Big(i\gamma_1 .
\partial_1 -m_1 - ie^2\gamma_1^{\mu}\widetilde A_{\mu}(y_1,x_2,x_2')
\Big)\psi_1(y_1)\bigg\}>, 
\nonumber \\
& & \\
\label{e7}
\widetilde A_{\mu}(y_1,x_2,x_2') &=& \int_{x_2^{\prime 0}}^{x_2^0} dy_2^0
D_{\mu 0}(y_1^0-y_2^0,\bf{y}_1-\bf{x}_2).
\end{eqnarray}
\par

\newpage 

\section{Coulomb gauge calculation of the energy spectrum}
\setcounter{equation}{0}

In the Coulomb gauge, $\widetilde A_{\mu}$ is:
\begin{equation} \label{e8}
\widetilde A_{\mu}^{(C)}(y_1,x_2,x_2') = ig_{\mu 0} \frac{1}{4\pi}\frac{1}
{|\bf{y}_1-\bf{x}_2|}\theta(x_2^0-y_1^0)\theta(y_1-x_2^
{\prime 0}).
\end{equation}
\par
The Green's function (\ref{e7}) can be evaluated by considering the 
differential equations it satisfies. These are:
\begin{eqnarray}
\label{e9}
\bigg[i\gamma_{10}\partial_{10}+i\gamma_1^i\partial_{1i}-m_1-\gamma_{10}
V(r)\theta(x_2^0-x_1^0)\theta(x_1^0-x_2^{\prime 0}\bigg]
G(x_1,x_2,x_1',x_2')\ & &\nonumber\\
= G_{20}(x_2'-x_2)i\delta^4(x_1-x_1'), & &\\
\label{e10}
G(x_1,x_2,x_1',x_2')\bigg[-i\gamma_{10}\stackrel{\leftarrow}{\partial_{10}'}
-i\gamma_1^i\stackrel{\leftarrow}{\partial_{1i}'}-m_1 - \gamma_{10}
V(r')\theta(x_2^0-x_1^{\prime 0})\theta(x_1^{\prime 0}-x_2^{\prime 0})
\bigg]\ & &\nonumber \\
= G_{20}(x_2'-x_2)i\delta^4(x_1-x_1'), & &
\end{eqnarray}
where
\begin{equation} \label{e11}
V(r)=-\frac{\alpha}{r},\ \ \ \ \ \alpha =\frac{e^2}{4\pi},\ \ \ \ \ 
r=|\bf{x}_1-\bf{x}_2|,\ \ \ \ \ r'=|\bf{x}_1'-\bf{x}_2'|.
\end{equation}
\par
Before solving equations (\ref{e9})-(\ref{e10}) we shall introduce a
particular representation for the one-particle Green's function which will
be useful throughout the resolution procedure. Let us first consider
the free Green's function of particle 1. Defining in momentum space the
functions
\begin{equation} \label{e12}
h_0(\bf{p}_1)=\gamma_0(\gamma_ip_1^i+m_1),\ \ \ \ \ \ 
\sqrt{h_0^2} = \sqrt{m_1^2+\bf {p}_1^2},
\end{equation}
and the projectors
\begin{equation} \label{e13}
\Lambda_{0\pm}=\frac{1}{2}\Big(1\pm\frac{h_0}{\sqrt{h_0^2}}\Big),
\end{equation}
one can write the free Green's function in the form:
\begin{equation} \label{e14}
G_{10}=\frac{i}{p_{10}-h_0+i\epsilon}\Lambda_{0+}\gamma_0
+\frac{i}{p_{10}-h_0-i\epsilon}\Lambda_{0-}\gamma_0.
\end{equation}
It takes in $x$-space the form:
\begin{equation} \label{e15}
G_{10}(x_1-x_1')=e^{-ix_1^0h_0}\frac{1}{2}\Big(\epsilon(x_1^0-x_1^
{\prime 0}) + \frac{h_0}{\sqrt{h_0^2}}\Big)\delta^3(\bf{x}_1
-\bf{x}_1')\gamma_0 e^{ix_1^{\prime 0} h_0'},
\end{equation}
where now $h_0$ and $h_0'$ are the $x$-space expressions of the function
$h_0(\bf{p}_1)$:
\begin{equation} \label{e16}
h_0=-i\gamma_0\gamma^i\partial_{1i}+m_1\gamma_0,\ \ \ \ \ 
h_0'=-i\gamma_0\gamma^i\stackrel{\leftarrow}{\partial_{1i}'}+m_1\gamma_0.
\end{equation}
By replacing, through the closure relation, the delta-function above by a  
complete set of spinor wave functions and identifying the
arguments of the exponential functions with the energy eigenvalues of these
functions, one deduces that for $x_1^0>x_1^{\prime 0}$ they correspond to
the positive energy solutions of the free Dirac equation and for $x_1^0<
x_1^{\prime 0}$ to the the negative energy ones, thus exhausting the 
spectrum of states saturating the free Green's function.
\par
The above construction can also be generalized to the case of the 
one-particle Green's function in the presence of a static vector potential 
$V(r)$. The Green's function then satisfies the differential equation
\begin{equation} \label{e17}
\Big(i\gamma_1 .\partial_1 - m_1 - \gamma_{10} V(r)\Big)G_1(x_1,x_1',
\bf{x}_2) = i\delta^4(x_1-x_1')
\end{equation}
and its conjugate one. Since $V$ is static, the formal solution of these
equations can be obtained from expression (\ref{e15}) by a shift of
the operators $h_0$ and $h_0'$ by the factor $V$. Defining:
\begin{equation} \label{e18}
h=-i\gamma_0\gamma^i\partial_{1i}+m_1\gamma_0+V(r), \ \ \ \ 
h'=-i\gamma_0\gamma^i\stackrel{\leftarrow}{\partial_{1i}'}+m_1\gamma_0
+V(r'),
\end{equation}
one obtains:
\begin{equation} \label{e19}
G_1(x_1,x_1',\bf{x}_2) = e^{-ix_1^0h}\frac{1}{2}\Big(\epsilon
(x_1^0-x_1^{\prime 0})+\frac{h}{\sqrt{h^2}}\Big)\delta^3(\bf{x}_1-
\bf{x}_1')\gamma_0 e^{ix_1^{\prime 0}h'}.
\end{equation}
[$\sqrt{h^2}$ is an appropriate generalization of $\sqrt{h_0^2}$.]
We also notice the following property of the operators $h$ and $h'$:
\begin{equation} \label{e20}
h\delta^3(\bf{x}_1-\bf{x}_1')\gamma_0=\delta^3(\bf{x}_1-
\bf{x}_1')\gamma_0 h',
\end{equation}
which allows us to convert, when necessary for symmetry reasons, the 
operators $h$ and $h'$ into each other. Again the use of the
closure relation through the delta-function allows us to identify
the spectrum os states saturating the Green's function with the set
of solutions of the Dirac equation in the presence of the static potential.
\par
We are now in a position to construct the solutions of Eqs. 
(\ref{e9})-(\ref{e10}). We first rewrite them in the form:
\begin{eqnarray} 
\label{e21}
\bigg[i\partial_{10}-h + V\Big(1-\theta(x_2^0-x_1^0)\theta(x_1-x_2^
{\prime 0})\Big) \bigg]G &=& G_{20}i\delta^4(x_1-x_1')\gamma_{10},\\
\label{e22}
G\bigg[-i\stackrel{\leftarrow}{\partial_{10}'}-h'+V|\Big(1-\theta(x_2^0
-x_1^{\prime 0})\theta(x_1^{\prime 0}-x_2^{\prime 0})\Big)\bigg] &=& 
G_{20} i\delta^4(x_1-x_1')\gamma_{10}.
\end{eqnarray}
After making the change of function $G=e^{-ix_1^0 h}\widetilde G
e^{ix_1^{\prime 0}h'}$, one integrates for $\widetilde G$, by imposing
the boundary condition that in the limit $x_2^{\prime 0}\rightarrow 
-\infty$ and $x_2^0\rightarrow \infty$, with $x_1^{\prime 0}$
and $x_1^0$ fixed, $G$ tends, up to multiplicative factors, to the static 
potential case solution (\ref{e19}), as is evident from Eqs. 
(\ref{e9})-(\ref{e10}). One finds the solution:
\begin{eqnarray} \label{e23}
G &=& G_{20}(x_2'-x_2) e^{-ix_1^0h} T \exp\Big\{i\int_{x_1^{\prime 0}}
^{x_1^0} dy_1^0 e^{iy_1^0h}
V(r)(1-\theta(x_2^0-y_1^0)\theta(y_1^0-x_2^{\prime 0}))e^{-iy_1^0h}\Big\}
\nonumber \\
& &\ \ \ \ \ \times \frac{1}{2}\Big(\epsilon(x_1^0-x_1^{\prime 0})
+\frac{h}{\sqrt{h^2}}\Big)\delta^3(\bf{x}_1-\bf{x}_1')\gamma_{10}
e^{ix_1^{\prime 0}h'},
\end{eqnarray}
where $T$ designates the chronological product. This expression can 
be further simplified for the situation that interests us. Sticking for
definiteness to the case $x_2^0$ and $x_1^0$ greater than $x_2^{\prime 0}$
and $x_1^{\prime 0}$, one observes that the domain of integration 
$[x_2^{\prime 0},x_2^0]$ does not contribute to the above integral. 
One therefore obtains two disjoint
domains of integrations, which allow the factorization of the $T$-product.
Using then the property (\ref{e20}), the expression of $G$ becomes:
\begin{eqnarray} \label{e25}
G &=& G_{20}(x_2'-x_2) e^{-ix_1^0h} T \exp\Big\{i\int_{x_2^0}^{x_1^0}
dy_1^0e^{iy_1^0h}V(r)(1-\theta(x_2^0-y_1^0)\theta(y_1^0-x_2^{\prime 0}))
e^{-iy_1^0h}\Big\}\nonumber \\
& &\ \ \ \ \ \ \times \frac{1}{2}\Big(\epsilon(x_1^0-x_1^{\prime 0})
+ \frac{h}{\sqrt{h^2}}\Big)\delta^3(\bf{x}_1-\bf{x}_1')\gamma_{10}
\nonumber \\
& &\ \ \ \ \times T \exp\Big\{i\int_{x_1^{\prime 0}}^{x_2^{\prime 0}} 
dy_1^0 e^{iy_1^0h'}V(r')(1-\theta(x_2^0-y_1^0)\theta(y_1^0-x_2^{\prime 0}))
e^{-iy_1^0h'}\Big\}.
\end{eqnarray}
Introducing the relative variables $x=x_1-x_2$ and $x'=x_1'-x_2'$, and
making in the integrals appropriate changes of variables, and also using
the property $\exp(e^BAe^{-B})=e^Be^Ae^{-B}$, one finally obtains for $G$
the expression:
\begin{eqnarray} \label{e26}
G &=& G_{20}(x_2'-x_2) T \exp\Big\{i\int_0^{x^0} dz^0 e^{-iz^0h}V(r)
\theta(x^0-z^0)e^{iz^0h}\Big\}\nonumber \\
& & \ \ \ \times e^{-ix_1^0h}\frac{1}{2}\Big(\epsilon(x_1^0-x_1^{\prime 0})
+\frac{h}{\sqrt{h^2}}\Big)\delta^3(\bf{x}_1-\bf{x}_1')\gamma_{10}
e^{ix_1^{\prime 0}h'}\nonumber \\
& & \ \ \ \times T \exp\Big\{-i\int_0^{x^{\prime 0}} dz^0 e^{-iz^0h'}
V(r')\theta(z^0-x^{\prime 0})e^{iz^0h'}\Big\}.
\end{eqnarray}
\par
Using for the delta-function the closure relation:
\begin{equation} \label{e27}
\delta^3(\bf{x}_1-\bf{x}_1') = \sum_n \psi_n(\bf{x}) \overline
\psi_n(\bf{x}')
\end{equation}
[from the expression of $G_{20}$, Eq. (\ref{e4}), we have $\bf{x}_2
=\bf{x}_2'$], where the $\psi_n$'s form a complete set of spinor wave
functions, and comparing the resulting expression of $G$ with its cluster
decomposition formula (for $x_2^0$ and $x_1^0$ greater than 
$x_2^{\prime 0}$ and $x_1^{\prime 0}$) \cite{gml}:
\begin{equation}  \label{e28}
G(x_1,x_2,x_1',x_2') =  \sum _n \Phi_n(x_1,x_2) \overline \Phi_n(x_1',x_2')\
=\ \sum _n \phi_n(x)e^{-iP_n.(X-X')}\ \overline \phi_n(x'), 
\end{equation}
where the $\Phi$'s are generalized Bethe--Salpeter wave functions, one deduces 
the energy spectrum of the theory. [We have introduced the total momentum 
variable $P=p_1+p_2$ and the total coordinates $X=(x_1+x_2)/2$ and $X'$.]
By identifying in the expression of $G$ the argument of the exponential
functions containing the total coordinates $X^0$ and $X^{\prime 0}$ with 
the energy eigenvalues, one deduces that the energy spectrum is that of the
positive energy solutions of the Dirac equation with the static potential 
$V$ (added by $m_2$):
\begin{equation} \label{e29}
\Big(i\gamma_1.\partial_1 -m_1-\gamma_{10}V(r)\Big) \psi(x_1^0,
\bf{x}_1-\bf{x}_2)=0.
\end{equation}
Repeating the above operations with values of $x_2^0$ and $x_1^
{\prime 0}$ greater than $x_2^{\prime 0}$ and $x_1^0$, one recovers the
negative energy solutions of Eq. (\ref{e29}). Therefore, the energy 
spectrum of the two-particle Green's function (\ref{e2}) in the 
limit $m_2\rightarrow \infty$ is given, up to the additive factor 
$m_2$, by the spectrum of eigenvalues of Eq. (\ref{e29}). This spectrum 
does not contain any abnormal solutions corresponding to the relative
time excitations.
\par
The previous procedure makes also possible the identification of the 
Bethe--Salpeter wave function in terms of the Dirac wave function
$\psi$. Designating by $p_{10}$ the energy eigenvalue of the Dirac wave 
function, one has for the bound state wave functions (for which $x_1^0>
x_1^{\prime 0}$):
\begin{equation} \label{e30}
\Phi(x_1,x_2) = e^{-i(m_2x_2^0+p_{10}x_1^0)} T \exp\Big\{i\int_0^{x^0}
dz^0 e^{-iz^0h}V(r)\theta(x^0-z^0)e^{iz^0h}\Big\} \psi(\bf{x}).
\end{equation}
For $x^0<0$, the integral vanishes and one obtains:
\begin{equation} \label{e31}
\Phi = e^{-i(m_2x_2^0+p_{10}x_1^0)}\psi(\bf{x}) = e^{-i(m_1+m_2+E)X^0}
e^{-i(m_1-m_2)x^0/2}e^{-iEx^0/2}\psi(\bf{x}),\ \ \ \ x^0<0,
\end{equation}
where we have introduced the binding energy $E$:
\begin{equation} \label{e32}
p_{10}=m_1+E,\ \ \ \ \ E<0.
\end{equation}
For $x^0>0$, the $\theta$-function in the integral can be replaced by 1,
and using the relation \cite{f} 
\begin{equation} \label{e33}
e^{(A+B)t}e^{-At}=T \exp\Big\{\int_0^tdt'e^{At'}Be^{-At'}\Big\},
\end{equation}
one finds:
\begin{eqnarray} \label{e34}
\Phi &=& e^{-i(m_2+p_{10})x_2^0} e^{-ix^0(h-V)}\psi(\bf{x})\nonumber \\
&=& e^{-i(m_1+m_2+E)X^0} e^{-i(m_1-m_2)x^0/2} e^{iEx^0/2} e^{-ix^0
(h_0-m_1)}\psi(\bf{x}),\ \ \ \ \ x^0>0.
\end{eqnarray}
In the bound state domain, the operator $(h_0-m_1)$ has a positive 
spectrum since it is essentially equivalent to the kinetic energy 
operator. One can then check that the above Bethe--Salpeter wave function 
has the correct spectral properties \cite{w,n1}. For $x^0>0$, it has
singularities for positive values of the relative energy variable, and
for $x^0<0$, it has singularities for negative values of the relative
energy variable. Furthermore, by decomposing $\Phi$ as
\begin{equation} \label{e35}
\Phi = \theta(x^0)\Phi_+ + \theta(-x^0)\Phi_-,
\end{equation}
one verifies that the conjugate Bethe--Salpeter wave function $\overline
\Phi$, identified from the comparison of Eqs. (\ref{e26}) and (\ref{e28})
with the use of Eq. (\ref{e27}), satisfies the correct definition (with the
antichronological product):
\begin{equation} \label{e36}
\overline \Phi = \theta(-x^0)\Phi_+^{\dagger}\gamma_{10} + \theta(x^0)
\Phi_-^{\dagger}\gamma_{10}.
\end{equation}
\par
The Bethe--Salpeter wave function could also have been constructed from
the resolution of the equations it satisfies. By taking in Eq. (\ref{e9})
the limit $x_2^{\prime 0}\rightarrow -\infty$ and using the cluster
decomposition (\ref{e28}), one finds the equations:
\begin{eqnarray} \label{e37}
& &\Big(i\gamma_1 .\partial_1 -m_1-\gamma_{10} V(r)\theta(x_2^0-x_1^0)
\Big)\Phi = 0, \nonumber \\
& & i(\partial_{10}+\partial_{20})\Phi = P_0\Phi,
\end{eqnarray}
the solution of which is precisely the function defined in Eqs. 
(\ref{e31}) and (\ref{e34}), when the continuity condition at $x^0=0$
is used.
\par

\newpage 

\section{Covariant gauges}
\setcounter{equation}{0}

For completeness, we briefly sketch the derivation in the case of linear
covariant gauges. Here, a complication arises because of the presence of
gauge photons (scalar and longitudinal) which create in the spectrum of
states additional cuts starting from the bound state poles \cite{jz}. These
gauge dependent cuts are usually removed by appropriately associating,
through Ward-Takahashi identities, with the exchanged photons the photons
contributing to the radiative corrections and taking in the cluster
decomposition formula (\ref{e28}) the limits $X^0\rightarrow \infty$ and
$X^{\prime 0}\rightarrow -\infty$, while keeping $x^0$ and $x^{\prime 0}$
finite. With this procedure one selects the bound state spectrum \cite{gml},
which is the object of interest. An illustration of this cancellation can
be found in Ref. \cite{js} for the static case. We shall not consider here
the radiative corrections, which are more complicated than in the static
limit, but shall indicate when necessary, their precise role.
\par
In linear covariant gauges, characterized by a gauge parameter $\xi$, 
the photon propagator in momentum space is:
\begin{equation} \label{e38}
D_{\mu\nu}(k) = -(g_{\mu\nu}-\xi\frac{k_{\mu}k_{\nu}}{k^2})\frac{i}
{k^2+i\epsilon}.
\end{equation}
Defining the functions
\begin{eqnarray}
\label{e39}
\chi'(x)&=& -\frac{ie^2}{4\pi^2}\frac{1}{2|\bf{x}|}\ln\left(\frac
{|\bf{x}|+x^0+i\epsilon}
{|\bf{x}|-x^0+i\epsilon}\right),\\
\label{e40}
\Delta(x) &=& \int\frac{d^4k}{(2\pi)^4} \frac{e^{-ik.x}}{(k^2+i\epsilon)^2}
= \frac{i}{16\pi^2}\ln(-\mu^2x^2),
\end{eqnarray}
[$\chi'$ being the derivative, in the Feynman gauge, with respect to $x^0$,
of the function $\chi$ defined in Ref. \cite{js},] one finds for the
effective potential $\widetilde A_{\mu}$ [Eq. (\ref{e7})] the expression:
\begin{equation} \label{e41}
e^2\widetilde A_{\mu}^{(\xi)}(x,x_2,x_2')=ig_{\mu 0}\Big(\chi'(x-x_2)-
\chi'(x-x_2')\Big) -ie^2\xi \partial_{\mu}\Big(\Delta(x-x_2)-\Delta
(x-x_2')\Big).
\end{equation}
The fact that the $\xi$-dependent part od $\widetilde A_{\mu,\xi}$ is a
derivative term, makes it possible its immediate elimination from Eqs. 
(\ref{e9})-(\ref{e10}). One finds for $G^{(\xi)}$ the following $\xi$
dependence:
\begin{equation} \label{e42}
G^{(\xi)} =
\exp\Big\{ie^2\xi\Big(\Delta(x_1-x_2)-\Delta(x_1-x_2')-\Delta(x_1'-x_2)
+\Delta(x_1'-x_2')\Big)\Big\} G^{(0)},
\end{equation}
which is precisely the general transformation law of the two-particle
Green's function under covariant gauge transformations of the photon
propagator (radiative corrections being neglected here) \cite{lk,z,o,bcm}.
It has been shown that these $\xi$-dependent gauge factors do not change
the bound state pole positions of Green's functions \cite{by,ffh}. It is
therefore sufficient to work in the Feynman gauge ($\xi=0$), which is also
similar to the scalar interaction case. (We shall henceforth omit the index
0 from the Green's function.)
\par
The function $\chi'$ has the following asymptotic behavior:
\begin{equation} \label{e43}
\chi'(x)_{\stackrel{{\displaystyle \longrightarrow}}{|x^0|\rightarrow
\infty}} \epsilon(x^0)\frac{1}{2}V(|\bf{x}|) + O(\frac{1}{x^0}).
\end{equation}
It is the $O(1/x^0)$ term which is gauge dependent (it is
proportional to the derivative of the function $\Delta$ of Eq. 
(\ref{e40}) and vanishes in the gauge $\xi=-2$) and must be
cancelled by a corresponding piece coming the radiative corrections 
\cite{js}. We shall assume that this cancellation has occurred and thus
the next-to-leading term in the asymptotic part of $\chi'$ is
$O(1/x^{02})$.
\par
Replacing $\widetilde A_{\mu}^{(0)}$ by its expression (\ref{e41}) in Eqs.
(\ref{e9})-(\ref{e10}) and isolating again the asymptotic part of the
potential, as in Eqs. (\ref{e21})-(\ref{e22}), one obtains:
\begin{eqnarray} \label{e44}
& &\left(i\gamma_1 .\partial_1-m_1-\gamma_{10}V(r)-\gamma_{10}\Big[\Big
(\chi' (x_2-x_1)-\frac {1}{2}V(r)\Big)+\Big(\chi'(x_1-x_2')-\frac{1}{2}
V(r)\Big) \Big]\right)G \nonumber \\
& &\ \ \ \ \ \ \ \ \ \ \ = G_{20}(x_2'-x_2)i\delta^4(x_1-x_1'),
\end{eqnarray}
and a similar conjugate equation. The remaining part of the resolution
of these equations parallels that utilized in the Coulomb gauge. One arrives
at an expression analogous to that of Eq. (\ref{e23}), where the 
potential part $V(1-\theta\theta)$ is replaced by the bracket term of 
Eq. (\ref{e44}) (with $x_1^0$ replaced by the integration variable $y_1^0$).
One then takes the limits $x_2^0,\ x_1^0\rightarrow \infty$ and $x_2^{\prime
0},\ x_1^{\prime 0}\rightarrow -\infty$, by keeping $x^0$ and $x^{\prime 0}$
finite. Because the difference $\chi'(x)-V/2$ behaves as $O(1/x^{02})$,
in the part containing $\chi'(x_2-y_1)-V/2$ it is the values of $y_1^0$
close to $x_2^0$ that contribute, while in the part containing
$\chi'(y_1-x_2')-V/2$ it is the values of $y_1^0$ close to $x_2^{\prime 0}$
that are relevant. Thus these two parts receive contributions from disjoint
intervals and the $T$-product factorizes again [Eq. (\ref{e25})], with
the first integral lying from $-\infty$ to $x_1^0$ and the second 
integral from $x_1^{\prime 0}$ to $\infty$. One ends up, for the bound
state spectrum, with the same conclusion as in the Coulomb gauge.
\par

\end{document}